\documentclass[aps,prd,showpacs,preprint,amssymb,nofootinbib]{revtex4}

\usepackage{amssymb,amsmath}
\usepackage{amsfonts}
\usepackage{mathrsfs}
\usepackage{txfonts}
\usepackage{amssymb}

\begin{document}

%%%%%%%%%%%%%%%%%%%%%%%%%%%%%  TITLE   %%%%%%%%%%%%%%%%%%%%%%%%%%%

\title[SitSit]{Massless scalar field in de Sitter spacetime: unitary quantum time evolution}

\author{Jer\'onimo Cortez}
\affiliation{Departamento de F\'\i sica,
Facultad de Ciencias, Universidad Nacional Aut\'onoma de
M\'exico, M\'exico D.F. 04510, Mexico.}
\email{jacq@ciencias.unam.mx}
\author{Daniel Mart\'in-de~Blas, Guillermo A. Mena
Marug\'an}
\affiliation{Instituto de Estructura de la Materia, IEM-CSIC,
Serrano 121, 28006 Madrid, Spain.}
\email{daniel.martin@iem.cfmac.csic.es, mena@iem.cfmac.csic.es}
\author{Jos\'e M. Velhinho}
\affiliation{Departamento de F\'{\i}sica, Faculdade de Ci\^encias, Universidade
da Beira Interior, R. Marqu\^es D'\'Avila e Bolama,
6201-001 Covilh\~a, Portugal.}
\email{jvelhi@ubi.pt}

\begin{abstract}

We prove that, under the standard conformal scaling,
a free scalar field in de Sitter spacetime admits an $O(4)$-invariant Fock quantization such that time evolution is unitarily implemented. Since this applies in particular to the massless case, this result disproves previous claims in the literature. We discuss the relationship between this quantization with unitary dynamics and the family of $O(4)$-invariant Hadamard states given by Allen and Folacci, as well as with the Bunch-Davies vacuum.
\end{abstract}


\maketitle

\section{Introduction}
\label{sec:intro}

In recent years, a series of remarkable results have been proven concerning the canonical quantization of linear scalar fields propagating on compact spatial manifolds, with a dynamics resembling that of a free field, but with an effective time-dependent mass. More precisely, the results apply to a field equation of the type
\begin{equation}
\label{s1}
\ddot\chi-\Delta\chi+s(t)\chi=0,
\end{equation}
defined on a static spacetime of the form $\Sigma\times I$, where $I$ is an interval of the real line, $\Sigma$ is a compact Riemannian manifold of dimension $d\leq 3$, $\Delta$ is the Laplace-Beltrami (LB) operator on $\Sigma$, and $s(t)$ is a general function of time, subject only to rather mild conditions. For instance, for all purposes, it suffices that this function is twice differentiable, with a second derivative that is integrable in each compact subinterval of $I$.

The results of Refs. \cite{PRD79,CMV8,CMOV-FTC} show that one can always find a Fock quantization for the system such that: i) the state that defines the Fock representation\footnote{This state is usually called the vacuum of the representation, although it does not necessarily correspond to an eigenstate of a Hamiltonian operator.} is invariant under the isometries\footnote{More generally, one can consider the group of  transformations that commute with the LB operator and are unitary in the Hilbert space of square integrable functions of the field configuration, with respect to the measure defined by the metric volume element.} of the spatial manifold $\Sigma$, and ii) the dynamics dictated by the field equation \eqref{s1} is unitarily implemented.

We notice that, although linear, the classical dynamics is nontrivial, owing to the presence of the time-dependent mass $s(t)$. By unitary implementation of the dynamics we mean the following: given two arbitrary values of time, $t$ and $t'$, the linear symplectic transformation that corresponds to classical canonical evolution from $t$ to $t'$ is implemented in the Fock representation as a unitary operator. This is assumed to happen for all instants of time in (every connected component of) the interval $I$, with no further conditions. In particular, no continuity conditions are imposed, and therefore nothing is said about the existence of a well-defined Hamiltonian operator.

In addition, and most importantly, the above Fock representation has been shown to be unique modulo unitary equivalence, in the sense that any other Fock representation with the same properties of invariance under spatial symmetries {\emph{and}} unitary dynamics is guaranteed to define a unitarily equivalent representation  \cite{PRD79,CMV8,CMOV-FTC}. Note that these uniqueness results are obtained in the absence of time-translation invariance, which is a key ingredient in the standard uniqueness theorems, regarding the quantization of free fields in stationary (or static) spacetimes (see Refs. \cite{kay,wald,ash-magnon}). In the considered nonstationary settings, the mentioned new results single out a unique equivalence class of representations, ensuring the nonambiguity of the physical predictions, rather than selecting a definite Fock representation based on a specific vacuum.

Though rather simple at first glance, these results find application in a variety of situations, including the quantization of inhomogeneous spacetimes, such as the Gowdy models (see Refs. \cite{unit-gt3,unique-gowdy-1,scho,BVV2,CQG25,hybrid-gowdy}), the quantization of cosmological perturbations \cite{CMV8,PRD85,PRD86}, and the discussion of string dynamics in arbitrary plane wave backgrounds \cite{string}. Another particular instance in which these results can be applied is the case of free fields in a nonstationary spacetime which is nevertheless conformal to a static universe, by means of a time dependent conformal factor. This is the situation found in Friedmann-Robertson-Walker (FRW) universes (with compact spatial sections), as well as in the very interesting case of the de Sitter spacetime\footnote{For other criteria concerning free fields in 1+1 dimensional de Sitter spacetime, see Ref. \cite{jac}.}. In fact, in all of these examples, the use of conformal time combined with a suitable scaling of the original field variable $\phi$ transforms the
original free field equation into and equation of the type \eqref{s1}. The scaled field that satisfies this latter equation is $\chi=\Omega^{1/2}\phi$, where $\Omega$ is the (exclusively) time-dependent conformal factor. The new field possesses an effective time-dependent mass $s(t)$ that depends on the conformal factor and on the mass of the original field.

However, one can find claims in the recent literature stating that, for the case of the massless field in de Sitter spacetime, no scaling of the field variable allows for a unitary dynamics. That is the conclusion of Ref. \cite{VV}. One of the aims of the present work is to clarify this situation, showing with due care that those claims are unsound\footnote{The reasons can be traced back to an unsuitable choice of momentum field, as well as to the use of arguments based on the limit of infinite times, which is radically different from the limit in which the number of modes grows to infinity (for further details, see Sec. \ref{sec:unitary dynamics}).}. The standard conformal scaling of the field does indeed lead to a field formulation with the desired properties, allowing for a representation where one can reach a unitary implementation of the dynamics of the scaled field $\chi$, regardless of the value of the mass of the original field in de Sitter spacetime. In fact, in what concerns the possibility of a unitary implementation of the dynamics, the value of the mass parameter, positive, null, or even negative, is not relevant. The value of the mass {\em is} of course important for the existence of a fully de Sitter invariant Fock state of the Hadamard
type. Actually, it is well known that the so called Bunch-Davies vacuum (or Euclidean vacuum), which is an $O(1,4)$-invariant state, breaks down in the massless case \cite{Allen}. Nonetheless, this does not prevent a consistent Fock quantization from being obtained, e.g. by means of $O(4)$, rather than $O(1,4)$, invariant states. We show explicitly that a Fock quantization can be achieved such that the dynamics of the massless field is unitarily implemented at the quantum level. This result is in full agreement with more general mathematical-physics results derived in the context of fields with a time-dependent mass. Taking advantage of such more general studies, one can moreover show that the obtained quantization is in fact unique, in the sense that any other $O(4)$-invariant Fock representation which also allows for a unitary dynamics is necessarily unitarily equivalent. In particular, the Fock representation that naturally emerges from our approach is seen to be unitarily equivalent to the $O(4)$-
invariant quantizations proposed some time ago by Allen and Folacci \cite{AllenFol}. Moreover, by applying our approach to the free massive field case, we get a Fock representation which (i) admits a unitary implementation of the dynamics and (ii) has a vacuum state which is unitarily equivalent to the celebrated Bunch-Davies state.

The paper is organized as follows. In Sec. \ref{sec:model} we introduce the classical setting for free, real scalar fields propagating in de Sitter spacetime. Conformal time is chosen as the evolution parameter and a scaling of the original field variable by the spacetime conformal factor is performed to define the basic field variable. The issue of unitary dynamics for the massless field is discussed in Sec. \ref{sec:unitary dynamics}. As the main result of the present work, a Fock representation permitting a unitary implementation of  the dynamics of the (scaled) massless scalar field is presented. Section \ref{sec:hadamard equivalence} is devoted to the study of the relationship between the so obtained Fock representation and those based on Hadamard states, namely the Bunch-Davies vacuum and the Allen-Folacci states. The relationship with another important set of states, namely the adiabatic states, is briefly addressed in Appendix \ref{A1:Adiab-states}. We  summarize and discuss our conclusions in Sec. \
ref{section: conclusions}. Finally, Appendix \ref{A2:beta-coef} contains details of the calculations needed in the proofs of the unitary evolution and the unitary equivalence between different vacua.

\section{The model: free fields in de Sitter spacetime}
\label{sec:model}

The de Sitter spacetime is the maximally symmetric spacetime of positive constant curvature. It has the topology of $\mathbb{R}\times S^3$ and can be seen as the hyperboloid
\begin{equation}
-x_0^2+ x_1^2+ x_2^2+ x_3^2+x_4^2= H^{-2}
\end{equation}
embedded in flat five-dimensional spacetime (see e.g. Refs. \cite{AllenFol,HawkEll}). The curvature of the space is $r=12 H^2$.

A system of coordinates $(t,\sigma,\theta,\varphi)$ in the whole space can be defined as follows:
\begin{eqnarray}
 & x_0 & = H^{-1} \sinh(Ht), \quad -\infty < t <\infty ,\\
 & x_1 & = H^{-1} \cosh(Ht)\cos(\sigma), \quad  0\leq \sigma\leq\pi ,\\
 & x_2 & = H^{-1} \cosh(Ht)\sin(\sigma)\cos(\theta), \quad  0\leq\theta\leq\pi ,\\
 & x_3 & = H^{-1} \cosh(Ht)\sin(\sigma)\sin(\theta)\cos(\varphi), \quad 0 \leq \varphi\leq 2\pi ,\\
 & x_4 & = H^{-1} \cosh(Ht)\sin(\sigma)\sin(\theta)\sin(\varphi).
\end{eqnarray}
In these coordinates the metric reads
\begin{equation}
\label{propert}
ds^2= -dt^2+H^{-2}\cosh^2(Ht)\{d\sigma^2+\sin^2(\sigma)[d\theta^2+\sin^2(\theta)d\varphi^2]\}.
\end{equation}
The de Sitter spacetime is conformal to the static universe $\mathbb{R}\times S^3$. To see this, let us introduce the conformal time
\begin{equation}
\label{eta-t-relation}
\eta=2\arctan(e^{Ht}),\quad \eta\in(0,\pi).
\end{equation}
In the new coordinate system, the metric takes the form
\begin{equation}
ds^2= a^2(\eta)\{-d\eta^2+d\sigma^2+\sin^2(\sigma)[d\theta^2+\sin^2(\theta)d\varphi^2]\},
\end{equation}
where one recognizes the metric of the static universe and the time-dependent conformal factor
\begin{equation}
\label{scale}
a(\eta)=\frac{1}{H\sin{\eta}}.
\end{equation}

Let us consider the propagation of a free (minimally coupled) real scalar field $\phi$ with mass $m$ and dynamical equation
\begin{equation}
(\Box -m^2)\phi=0,
\end{equation}
where $\Box$ stands for the d'Alambertian associated to the spacetime metric. In the coordinates $(t,\sigma,\theta,\varphi)$, this field equation becomes
\begin{equation}
\label{infeq}
\partial^2_t\phi+3H\tanh(Ht)\partial_t\phi+\left(-\frac{\Delta}{a^2}+m^2\right)\phi=0.
\end{equation}
Here, $\Delta$ is the LB operator on $S^3$. We now change to conformal time and introduce the scaled field\footnote{Similar scalings can also be performed in less than three spatial dimensions to reach a field which admits a unitary dynamics \cite{CMOV-FTC}.}
\begin{equation}
\label{rescale}
\chi=a \phi.
\end{equation}
We then obtain the new field equation
\begin{equation}
\label{feq}
\ddot{\chi}
+\left[-\Delta +(m^2-2H^2){a^2}+1\right]\,\chi=0.
\end{equation}
The dot stands for the derivative with respect to the conformal time $\eta$.
Note the absence of terms containing the first time derivative of the field.

{The Lagrangian density corresponding to the field $\chi$ (up to total time derivatives) is
\begin{equation}
L=\frac{1}{2}\left[\left(\dot{\chi}\right)^2 -\left(\nabla\chi\right)^2 -\left(m^2 a^2-\frac{\ddot{a}}{a}\right)\,\chi^2\right].
\end{equation}
The canonical momentum conjugate to $\chi$ can then be taken as}
\begin{equation}
\label{momentum}
P_\chi= \dot{\chi}.
\end{equation}
Alternatively, one might have started from the canonical pair $(\phi,P_\phi=a^{3}\partial_t\,\phi)$ and performed the time-dependent canonical transformation
\begin{equation}
\label{tdct}
\chi  =  a\phi, \qquad P_\chi  =  \frac{P_\phi}{a}+\dot a \phi.
\end{equation}

Let us return to the field equation \eqref{feq} and decompose the field $\chi$ in eigenmodes of the LB operator:
\begin{equation}
\label{modes}
\chi=\sum_{k,\ell,m} q_{k\ell m} Y_{k\ell m},
\end{equation}
where $Y_{k\ell m}$ are the $S^3$-spherical harmonics, which satisfy the eigenvalue equation
\begin{equation}
\Delta Y_{k\ell m}=-k(k+2)Y_{k\ell m},
\end{equation}
and provide an orthonormal basis for the space of square integrable functions on $S^{3}$. In these formulas, the integer $k$ takes values from $0$ to $\infty$, $\ell$ varies from $0$ to $k$, and $m$ ranges from $-\ell$ to $\ell$ (see e.g. Refs. \cite{sphar1,sphar2}). In the following, we will use the notation $q_{\bf k}$ to collectively denote all the modes  $q_{k\ell m}$ corresponding to the same value of $k$.

Introducing the decomposition \eqref{modes} in the field equation \eqref{feq} and using the orthogonality relations for the harmonics, we obtain the dynamical equation of each mode:
\begin{equation}
\label{modeeq}
\ddot q_{\bf k}+\left[(k+1)^2 +\frac{m'^2-2}{\sin^{2}\eta}\right]q_{\bf k}=0,
\end{equation}
where $m'=m/H$. These equations are the same for all the values of $\ell$ and $m$ that share the same value of $k$. Note also that, from $P_\chi=\dot \chi$, it follows that the momentum canonically conjugate to the mode variable $q_{\bf k}$ satisfies the equation $p_{\bf k}=\dot q_{\bf k}$.

The general solution to the equations of motion \eqref{modeeq} is well known. In fact, a change of variable $q_{\bf k}(\eta)=(\sin\eta)^{1/2} f(-\cos\eta)$ transforms equation \eqref{modeeq} into
\begin{equation}
\label{legendre}
\left(1-y^2\right)\frac{d^2f}{dy^2}-
2y\frac{df}{dy}+\left[\left(k+1\right)^2-\frac{1}{4}-\frac{1}{1-y^2}\left(\frac{9}{4}-m'^2\right)\right]f(y)=0,
\end{equation}
with $y=-\cos\eta$. This is a Legendre equation of degree $\nu=k+1/2$ and order $\mu=\left(9/4-m'^2\right)^{1/2}$, whose independent solutions are the associated Legendre functions $P^\mu_\nu$ and $Q^\mu_\nu$ (see e.g. Refs. \cite{AS,GR}). The general solution to the equations of motion \eqref{modeeq} is therefore
\begin{equation}
\label{solu}
q_{\bf k}(\eta)=A_{\bf k}\sqrt{\sin\eta}P^\mu_\nu(-\cos\eta)+ B_{\bf k}\sqrt{\sin\eta}Q^\mu_\nu(-\cos\eta),
\end{equation}
where $A_{\bf k}$ and $B_{\bf k}$ are arbitrary (complex) constants.

In the canonical formalism, the general solution to the corresponding Hamiltonian equations
\begin{equation}
\dot q_{\bf k} = p_{\bf k}, \qquad \dot p_{\bf k} = -\left[(k+1)^2 +(m'^2-2)\sin^{-2}\eta\right]q_{\bf k},
\end{equation}
can then be written in the form
\begin{equation}
\label{defM}
\left(\begin{array}{c} q_{\bf k} \\ p_{\bf k} \end{array}\right)=
M_{k}(\eta)\left(\begin{array}{c} A_{\bf k} \\  B_{\bf k}\end{array}\right),\qquad
M_k(\eta)=\left(\begin{array}{cc} R^\mu_\nu(-\cos\eta)\:\:\: &  S^\mu_\nu(-\cos\eta)\\
 \dot{R}^{\mu}_{\nu}(-\cos \eta)\:\:\: & \dot{S}^{\mu}_{\nu}(-\cos\eta) \end{array}\right),
\end{equation}
where $R^\mu_\nu$ and $S^\mu_\nu$ are given by
\begin{equation}
\label{defRS}
R^{\mu}_{\nu}(-\cos\eta)=\sqrt{\sin\eta}P^{\mu}_{\nu}(-\cos\eta),
\qquad S^{\mu}_{\nu}(-\cos\eta)=\sqrt{\sin\eta}Q^{\mu}_{\nu}(-\cos\eta).
\end{equation}

By using the relation
\begin{equation}
\frac{dP^\mu_\nu(y)}{dy}= \frac{1}{y^2-1}[\nu y P^\mu_\nu(y)-(\nu +\mu)P^\mu_{\nu -1}(y)],
\end{equation}
valid also for $Q^\mu_\nu(y)$ (see e.g. Refs. \cite{AS,GR}), $\dot{R}^{\mu}_{\nu}(-\cos\eta)$ and $\dot{S}^{\mu}_{\nu}(-\cos\eta)$ can be expressed as follows
\begin{eqnarray}
\label{dotR}
\dot{R}^{\mu}_{\nu}(-\cos\eta) & = & \frac{1}{\sqrt{\sin\eta}} \left[\left(\nu+1/2\right)\cos\eta\, P^{\mu}_{\nu}(-\cos\eta)+(\nu+\mu)P^{\mu}_{\nu-1}(-\cos\eta)\right], \\
\label{dotS}
\dot{S}^{\mu}_{\nu}(-\cos\eta) & = & \frac{1}{\sqrt{\sin\eta}}\left[\left(\nu+1/2\right)\cos\eta\, Q^{\mu}_{\nu}(-\cos\eta)+(\nu+\mu)Q^{\mu}_{\nu-1}(-\cos\eta)\right].
\end{eqnarray}

Note that one can write the above matrix elements directly in terms of the argument $\cos\eta$, instead of $- \cos\eta$, since \cite{GR}
\begin{eqnarray}
\label{minus1}
P^\mu_\nu(-x)&=&\cos[(\nu+\mu)\pi]P^\mu_\nu(x)-\frac{2}{\pi}\sin[(\nu+\mu)\pi]Q^\mu_\nu(x),
\\
\label{minus2}
Q^\mu_\nu(-x)&=&-\cos[(\nu+\mu)\pi]Q^\mu_\nu(x)-\frac{\pi}{2}\sin[(\nu+\mu)\pi]P^\mu_\nu(x).
\end{eqnarray}
It is not difficult to see that the same expressions hold for $R^{\mu}_{\nu}$ and $S^{\mu}_{\nu}$ [replacing $P^{\mu}_{\nu}(\pm x)$ and $Q^{\mu}_{\nu}(\pm x)$ with $R^{\mu}_{\nu}(\pm x)$ and $S^{\mu}_{\nu}(\pm x)$, respectively].
In particular, for the massless case ($\mu=3/2$), we get simply
\begin{equation}
\label{pm}
R^\mu_\nu(-x)=(-1)^k R^\mu_\nu(x),\qquad S^\mu_\nu(-x)=(-1)^{k+1} S^\mu_\nu(x),
\end{equation}
where we have used that $\nu=k+1/2$.

Time evolution from an arbitrary reference time $\eta_0$ to any another time $\eta$ is then given by the canonical transformation
\begin{equation}
\label{evolution}
\left(\begin{array}{c} q_{\bf k}(\eta) \\ p_{\bf k}(\eta) \end{array}\right)=
U_{k}(\eta_0,\eta)\left(\begin{array}{c} q_{\bf k}(\eta_0) \\  p_{\bf k}(\eta_0)\end{array}\right),
\end{equation}
where  $U_{k}(\eta_0,\eta)=M_{k}(\eta)M^{-1}_{k}(\eta_0)$.

\section{Quantization with unitary dynamics}
\label{sec:unitary dynamics}

We will now show that one can find a Fock quantization which allows a unitary implementation of the dynamics of the field $\chi$, i.e., a unitary implementation of all transformations \eqref{evolution}, for arbitrary values of $\eta$  and $\eta_0$.

A Fock quantization is defined by a choice of a complex structure on phase space\footnote{Remember that a complex structure is a map on phase space that preserves the canonical structure and whose square is minus the identity. For the construction of a Fock representation, one demands that the complex and the canonical structures be compatible in the sense that a suitable composition of their actions be positive definite.}, which is tantamount to a choice of sets of creation and annihilation variables (up to irrelevant changes which do not mix both sets). Let us introduce the classical (complex) variables
\begin{equation}
\label{ca-var}
\displaystyle
a_{\bf k}={\frac{1}{\sqrt{2\omega_k}}} \left(\omega_k q_{\bf k} +i p_{\bf k}\right), \qquad a^*_{\bf k}={\frac{1}{\sqrt{2\omega_k}}} \left(\omega_k q_{\bf k} -i p_{\bf k}\right),
\end{equation}
where the frequency $\omega_k$ is chosen to match the time-independent part in the equations of motion \eqref{modeeq}, i.e. $\omega_k=k+1$. Naturally, these variables satisfy canonical Poisson brackets $\{a_{\bf k},a^*_{\bf k'}\}=i\delta_{\bf k,\bf k'}$.

If we now declare that the  variables $a_{\bf k}$ and $a^*_{\bf k}$ are to be quantized as creation and annihilation operators, then we single out a particular Fock quantization. In other words, the complex structure $J$ that determines our particular Fock quantization is defined by $J(a_{\bf k})=ia_{\bf k}$, $J(a^*_{\bf k})=-ia^*_{\bf k}$ (see Ref. \cite{CMV8} for details on the Fock quantization).

Classical time evolution, written in terms of the variables $a_{\bf k}$, $a^*_{\bf k}$, is given by
\begin{equation}
\label{evolution-a}
\left(\begin{array}{c} a_{\bf k}(\eta) \\ a^*_{\bf k}(\eta) \end{array}\right)=
{\cal U}_{k}(\eta_0,\eta)\left(\begin{array}{c} a_{\bf k}(\eta_0) \\  a^*_{\bf k}(\eta_0)\end{array}\right) =
T_{k}U_{k}(\eta_0,\eta)T^{-1}_k\left(\begin{array}{c} a_{\bf k}(\eta_0) \\  a^*_{\bf k}(\eta_0)\end{array}\right),
\end{equation}
where
\begin{equation}
\label{defT}
T_k={\frac{1}{\sqrt{2\omega_n}}}\left(
\begin{array}{cc}
\omega_n & i\\
\omega_n & -i\end{array}\right)
\end{equation}
is the matrix corresponding to the change of variables \eqref{ca-var}. Since the transformations \eqref{evolution-a} are canonical, they necessarily take the form of a general Bogoliubov transformation, i.e., they can be written in the form
\begin{eqnarray}
\label{bogo-transf} {\cal U}_{k}(\eta_{0},\eta) = \left( \begin{array}{cc}
\alpha_k(\eta_0,\eta) &\beta_k(\eta_0,\eta)\\
\beta_k^*(\eta_0,\eta) & \alpha_k^*(\eta_0,\eta)\end{array}\right),
\end{eqnarray}
with
\begin{equation}
|\alpha_k(\eta_0,\eta)|^2-|\beta_k(\eta_0,\eta)|^2=1,
\end{equation}
independently of the particular values of $k$, $\eta_0$, and $\eta$.

Standard results \cite{shale,honegger} now show (see Ref. \cite{CMV8} for details) that the dynamics is unitarily implementable (in the above Fock quantization) if and only if the functions $\beta_k$ in Eq. \eqref{bogo-transf} are square summable; that is, if and only if
\begin{equation}
\label{cond}
\sum_{k=0}^\infty\sum_{\ell=0}^k\sum_{m=-\ell}^{\ell}|\beta_k(\eta_0,\eta)|^2=
\sum_{k=0}^\infty(k+1)^2|\beta_k(\eta_0,\eta)|^2<\infty,
\end{equation}
where the degeneracy factor $(k + 1)^2$ counts the number of degrees of freedom with the same dynamics. The fulfillment of this summability condition depends on the asymptotic behavior of the functions $\beta_k(\eta_0,\eta)$, for large values of $k$. This in turn depends on the asymptotics of $R^\mu_\nu$ and $S^\mu_\nu$ dictated by the ultraviolet (large values of the degree $\nu=k+1/2$) behavior of the Legendre functions $P^\mu_\nu$ and $Q^\mu_\nu$. The expansions of $P^\mu_\nu$ and $Q^\mu_\nu$ are as follows \cite{GR}
\begin{equation}
\label{seexR}
P^\mu_\nu(\cos\eta)=\sqrt{\frac{2}{\pi}}
\frac{\Gamma(\nu+\mu+1)}{\sqrt{\sin\eta}}\sum_{j=0}^{\infty}\frac{\Gamma(\mu+j+1/2)\,
\cos\left[\left(\nu+j+1/2\right)\eta+\frac{\pi}{4}(2j-1)+\frac{\mu \pi}{2}\right]}{\Gamma(\mu-j+1/2)\,\Gamma(j+1) \,\Gamma(\nu+j+3/2)\,(2\sin\eta)^{j}},
\end{equation}
\begin{equation}
\label{seexS}
Q^\mu_\nu(\cos\eta)=\sqrt{\frac{2}{\pi}}
\frac{\Gamma(\nu+\mu+1)}{\sqrt{\sin\eta}}\sum_{j=0}^{\infty}\frac{\Gamma(\mu+j+1/2)\,
\cos\left[\left(\nu+j+1/2\right)\eta-\frac{\pi}{4}(2j-1)+\frac{\mu \pi}{2}\right]}{(-1)^{j}\,\Gamma(\mu-j+1/2)\,\Gamma(j+1) \,\Gamma(\nu+j+3/2)\,(2\sin\eta)^{j}}.
\end{equation}
These formulas provide asymptotic expansions (up to arbitrary order) valid not only for real values of the parameter $\mu$, but also for complex ones. Thus, the range of validity of the asymptotic expansions covers the whole set of possible values of $m$, including, of course, the massless case ($\mu=3/2$).

Let us momentarily focus on the massless case. The asymptotic behavior of the Bogoliubov coefficients $\beta_k(\eta_0,\eta)$ can now be deduced, taking into account the matrices $M_k$, relations (\ref{dotR}, \ref{dotS}), the matrices $U_k$ and  $T_k$, identities (\ref{pm}), and the asymptotics for the functions $P^{3/2}_\nu$ and  $Q^{3/2}_\nu$. A lengthy but straightforward computation, detailed in Appendix \ref{A2:beta-coef}, shows that $\beta_k(\eta_0,\eta)$ is of order $O\left(k^{-2}\right)$, for large $k$, i.e., the asymptotic behavior when $k\to \infty$ is given by
\begin{equation}
\label{abeta}
\beta_k(\eta_0,\eta)= {O\left(k^{-2}\right)},\qquad \forall \eta,\eta_0.
\end{equation}
Thus, it follows that the summability condition \eqref{cond} is satisfied for all values of $\eta_0$ and $\eta$, and therefore the dynamics is unitarily implemented in the considered Fock representation\footnote{Since the  complex structure $J$ depends on the LB operator only, which in turn is an $O(4)$-invariant object, our unitary Fock representation is, in addition, an $O(4)$-invariant quantum theory.
Let us note also that the same asymptotic behavior and therefore the same conclusions
apply for any other value of the mass {(see Appendix \ref{A2:beta-coef})}.}. This result is in complete agreement with the general results proven in Refs. \cite{CMV8,CMOV-FTC}, and disproves the conclusion of Ref. \cite{VV}, where it is claimed that one cannot attain (by means of a Fock quantization) quantum unitarity of the evolution for the massless field in de Sitter spacetime, independently of the field redefinition $\phi\to f(t)\phi$.

{The calculations in Ref. \cite{VV} are based on a specific choice of momentum for the scaled field: the momentum obtained just by the inverse scaling. Nevertheless, one can also adopt other choices of momentum in order to obtain a canonical pair, while respecting the linearity of the system. Namely, one can allow for a time-dependent linear contribution of the field configuration to the momentum. This kind of time-dependent linear transformations affect the dynamics of the basic field variables. Actually, time-dependent canonical transformations of this type have been shown to be crucial to arrive at a unitary evolution \cite{Crit}. More specifically, a unique transformation is admissible when the dimension of the spatial hypersurfaces is greater than one, as it is the case here. The appropriate canonical transformation (including the field scaling) that leads to unitary evolution is that provided in Eq. \eqref{tdct}. In view of the transformation, it is particularly convenient to describe the system in conformal time, because then the privileged momentum is given just by the time derivative of the scaled field and the field equations simplify considerably, reflecting the conformal nature of the spacetime metric. In Ref. \cite{VV}, nonetheless, this conformal nature was not explored. Only in the concluding section of that work the conformal time was considered, presenting just a heuristic argument to support that the conclusions about the nonunitarity of the dynamics were valid as well for conformal time.} That argument is however not correct. The argument goes as follows. When the scaling \eqref{rescale} and the conformal time $\eta$ are used, the field equation that one obtains for the massless case, namely Eq. \eqref{feq} with $m=0$, is of the form of a  field with a time-dependent mass which is always negative. That is indeed the case, since we get from Eq. \eqref{feq}
\begin{equation}
\label{feqm0}
\ddot\chi-\Delta \chi+\left[1 -\frac{2}{\sin^2\eta}\right]\,\chi=0.
\end{equation}
The time-dependent mass $m(\eta)=1-2/\sin^2\eta$ is not only strictly negative, but moreover blows up when $\eta\to 0,\pi$ (which corresponds to $t\to\pm\infty$). It was then argued in Ref. \cite{VV} that, given that the time-dependent squared frequencies of the harmonic modes, namely $(k+1)^2-2/\sin^2\eta$, are all negative in the limit $t\to\pm\infty$, this would introduce a non-oscillatory behavior in that limit that would cause the failure of unitarity. However, the above argument does not really pose any obstruction to the unitary implementation of the dynamics. It rather points out that the ultraviolet limit, in which the infinite number of modes of the system come into the scene, and the limit of infinite time $t$ are radically different. In fact, unitary dynamics means the unitary implementation of all the evolution transformations \eqref{evolution} between any two {\em finite} values of time. Actually, when one considers the evolution between two instants of time, $\eta_0$ and $\eta$, one should
look to the values of $(k+1)^2-2/\sin^2\eta$ in the limit of large $k$, and not in the
limit of large $t$, because the dynamical transformation is sensitive only to the values in the interval $[\eta_0,\eta]$, since the equation of motion is local in time. What happens is that for all $k$ greater than some (maybe large, but finite) order $k_0$ (which depends on $\eta_0$ and $\eta$), the values of the squared frequencies  $(k+1)^2-2/\sin^2\eta$ are positive in all the  considered time interval. It is true that, for a finite number of modes, the evolution (between finite times) takes place with a negative time-dependent frequency. But this does not affect the possible unitary implementation of the dynamics in a fundamental way, because linear dynamics in finite dimensions is always unitary, as it is granted by the Stone-von Neumann uniqueness theorem \cite{Vonneu}.

For a full clarification of this situation, let us consider the case where the time dependent term $2/\sin^2\eta$ in the field equation is replaced simply by  $2/\eta^2$. This corresponds of course to the limit $t\to\pm\infty$ in the scale factor \eqref{scale} and is therefore actually physically relevant. Moreover, the field equation obtained with this replacement in the massless case, namely
\begin{equation}
\label{feqm0simpl}
\ddot\chi-\Delta \chi+\left[1-\frac{2}{\eta^2}\right]\chi=0,
\end{equation}
keeps precisely the qualitative features of the original equation \eqref{feqm0} that are involved in the argument sketched above about the behavior of the mode frequencies in the region of large $|t|$. The advantage of Eq. \eqref{feqm0simpl} is that it can be solved explicitly in terms of elementary functions, thus making the whole discussion fully transparent.

In fact, the equations of motion that we obtain for the harmonic modes, corresponding to Eq. \eqref{modeeq} (with $m=0$) are now
\begin{equation}
\label{newmodeeq}
\ddot q_{\bf k}+\left[(k+1)^2 -\frac{2}{\eta^2}\right]q_{\bf k}=0.
\end{equation}
One can readily check that the general solution is of the form
\begin{equation}
\label{newsolu}
q_{\bf k}(\eta)=A_{\bf k}\left(\frac{\sin[(k+1)\eta]}{(k+1)\eta}-\cos[(k+1)\eta]\right)+
B_{\bf k}\left(-\frac{\cos[(k+1)\eta]}{(k+1)\eta}-\sin[(k+1)\eta]\right),
\end{equation}
where $A_{\bf k}$ and $B_{\bf k}$ are arbitrary complex constants. One can now simply follow the procedure described above and obtain the corresponding coefficients $\beta_k(\eta_0,\eta)$ in the Bogoliubov transformation \eqref{bogo-transf}. With $\Delta\eta$ standing for $\eta-\eta_0$, we get in this case
\begin{eqnarray}
\beta_k(\eta_0,\eta)&=&\frac{1}{2(k+1)^2}\left(\frac{1}{\eta_0^2}-\frac{1}{\eta^2} +\frac{i\Delta\eta}{(k+1)\eta_0^2\eta^2}\right)\cos[(k+1)\Delta\eta] \nonumber \\
\,& & + \frac{i}{2(k+1)^2}\left(\frac{1}{\eta_0^2}+
\frac{1}{\eta^2}- \frac{1}{(k+1)^2\eta_0^2\eta^2}-\frac{i(\eta+\eta_0)}{(k+1)\eta_0^2\eta^2}\right)
\sin[(k+1)\Delta\eta].
\end{eqnarray}

The leading behavior is thus of order ${O\left(k^{-2}\right)}$, and it follows that the summability condition \eqref{cond} is indeed satisfied. This was the conclusion expected from more general results concerning the unitary implementation of time evolution \cite{CMV8,CMOV-FTC}, and confirms, in particular, that the sign of the mass term is irrelevant\footnote{A comment in this respect already appeared in Ref. \cite{string}.}.

We have therefore constructed an $O(4)$-invariant Fock representation for the massless free scalar field in de Sitter spacetime, with unitary dynamics. In what follows, we will investigate the relationship between our Fock vacuum and the family of $O(4)$-invariant Hadamard states characterized by Allen and Folacci \cite{AllenFol}. Before doing that, let us
stress that, although we have focused on the massless case, the unitarity result holds as well for any massive free field; i.e., the Fock representation, defined by the choice (\ref{ca-var}) of annihilation and creation operators, provides a quantum description where time evolution of the free massive field admits a unitary implementation. That this is so can be directly verified from the asymptotic behaviors of $P^{\mu}_{\nu}$ and $Q^{\mu}_{\nu}$ which, for any (constant) value of the parameter $\mu$ (including complex numbers), provide a beta coefficient satisfying Eq. (\ref{abeta}). Hence, given a scalar field with $m\geq 0$ (what is more, with any real
-even negative- value of $m$), there exists at least one Fock representation where time evolution is implemented as a unitary operator. Further details of this proof can be found in Appendix \ref{A2:beta-coef}.

\section{Equivalence with the Allen-Folacci's $O(4)$-invariant state}
\label{sec:hadamard equivalence}

An alternate way of defining a Fock quantization consists in selecting a particular set of complex mode solutions $\{u_{\bf k}\}$ to the equations of motion \eqref{modeeq}. These solutions are normalized so that they satisfy the condition
\begin{equation}
\label{decomp}
u_{\bf k}\dot{u}_{\bf k}^{\ast}-u^{\ast}_{\bf k}\dot{u}_{\bf k}=i
\end{equation}
on a given Cauchy surface, say $\eta=\eta_{0}$. The symbol $\ast$ denotes complex conjugation. Once such a set of solutions is chosen, one can write the general solution to the field equation \eqref{feq} as
\begin{equation}
\xi=\sum_{\bf k} \left(b_{\bf k}u_{\bf k}Y_{\bf k}+b^{\ast}_{\bf k}u^{\ast}_{\bf k}Y^{\ast}_{\bf k}\right).
\end{equation}
The Fock quantization is then performed by declaring that $b_{\bf k}$ and $b_{\bf k}^*$ are to be quantized as the annihilation and creation operators of the Fock representation (see Ref. \cite{LR} for details and a nice account on Fock states on homogeneous and isotropic spaces). In this description, the Fock quantization presented in the previous section corresponds to the choice of mode solutions determined by the following initial data:
\begin{equation}
\label{ourdata}
u_{\bf k}(\eta_0)=\frac{1}{\sqrt{2\omega_{k}}},\qquad \dot{u}_{\bf k}(\eta_0)=-i \sqrt{\frac{\omega_{k}}{2}}.
\end{equation}

For massive free fields in de Sitter spacetime, the  Fock quantization is usually carried out using the unique solution which is de Sitter invariant, i.e., invariant under the full $O(1,4)$ group, and satisfies the Hadamard criterion. The corresponding set of mode solutions is of the form \eqref{solu}, with \cite{Allen,AllenFol}
\begin{equation}
\label{ABeucvac}
B_{\bf k}=-\frac{2}{\pi} iA_{\bf k},\qquad A_{\bf k}=\sqrt{\frac{\pi}{4}\frac{\Gamma(k-\mu+3/2)}{\Gamma(k+\mu+3/2)}}e^{i\pi\mu/2}.
\end{equation}

The vacuum of the corresponding Fock representation is known as the Bunch-Davies, or Euclidean vacuum. Explicitly, the mode solutions determining this Euclidean quantization can be written as
\begin{equation}
\label{BD}
\chi_{\bf k}(\eta)=A_{\bf k}\left[R^\mu_\nu(-\cos\eta)- \frac{2}{\pi} i S^\mu_\nu(-\cos\eta)\right]=
A_{\bf k}e^{i(\nu+\mu)\pi}\left[R^\mu_\nu(\cos\eta)+ \frac{2}{\pi} i S^\mu_\nu(\cos\eta)\right],
\end{equation}
where in the last equality we have taken into account Eq. \eqref{defRS} as well as relations (\ref{minus1},\ref{minus2}). Given a Cauchy surface, say $\eta=\eta_0$, the momentum canonically conjugate to $\chi_{\bf k}(\eta_0)$ is
\begin{equation}
\label{dchi}
\dot{\chi}_{\bf k}(\eta_0)=
A_{\bf k}e^{i(\nu+\mu)\pi}\left.\left[\dot{R}^\mu_\nu(\cos\eta)+ \frac{2}{\pi} i\dot{S}^\mu_\nu(\cos\eta)\right]\right\vert_{\eta_0}.
\end{equation}

In the case of zero mass, it is a well known fact that the Euclidean vacuum breaks down, i.e., there is no de Sitter invariant Hadamard vacuum in this case \cite{Allen}. As explained in Refs. \cite{Allen,AllenFol}, this is due to the dynamics of the zero mode only. One can in fact easily check that taking $m=0$, and hence $\mu=3/2$, in the above expressions (\ref{ABeucvac},\ref{BD}), one gets perfectly well-defined solutions for $\bf k\not =0$, whereas the corresponding expression for $\bf k =0$ becomes meaningless. To obtain a complete set of well-defined solutions, and therefore a well-defined quantization, one only needs to derive proper solutions for the zero mode [or just quantize the zero mode in a different, consistent alternate way]. Independent solutions to the zero mode equation of motion with $m=0$ are $1/\sin\eta$ and $(\eta/\sin\eta) -\cos\eta$. It is shown in Ref.  \cite{AllenFol} that one can arrive at a one-parameter family of solutions for the zero mode such that [together with the solutions
\eqref{BD} for $\bf k\not=0$] $O(4)$-invariant Hadamard vacua are obtained.

The question naturally arises of whether or not the quantization presented in the previous section is unitarily equivalent to those corresponding to the $O(4)$-invariant states constructed by Allen and Folacci. Note that the unitary equivalence between those representations depends only on the behavior of the states for large values of $k$, and therefore the particular quantization used for the zero mode is irrelevant (provided of course that we are using for the zero mode a standard quantization that satisfies the Stone-von Neumann conditions, and not e.g. a polymer type quantization).

Given two sets of mode solutions $\{\chi_{\bf k}\}$ and $\{\chi'_{\bf k}\}$, determined by initial conditions $\{(\chi_{\bf k},\dot{\chi}_{\bf k}\}|_{\eta_{0}}$ and $\{(\chi'_{\bf k},\dot{\chi}'_{\bf k}\}|_{\eta_{0}}$, the two corresponding Fock representations are unitarily equivalent if and only if the following set of $\bar{\beta}_{\bf k}$ coefficients is square summable (see e.g. Ref. \cite{LR}):
\begin{equation}
\label{beta}
\bar{\beta}_{\bf k}=i\left[\dot{\chi}'_{\bf k}(\eta_0)\chi_{\bf k}(\eta_0)-
\dot{\chi}_{\bf k}(\eta_0)\chi'_{\bf k}(\eta_0)\right].
\end{equation}
Again, for the sets of solutions that we are considering, the coefficients $\bar{\beta}_{\bf k}$ depend only on the index $k$, and not on the full set of labels ${\bf k}$. So, the summability condition is still of the type (\ref{cond}), with a degeneracy factor $(k + 1)^2$.

Let us now fix a Cauchy surface, $\eta=\eta_0$, and evaluate the coefficients $\bar{\beta}_{\bf k}$ relating our data/representation \eqref{ourdata} with the Allen-Folacci data/representation $(\chi_{\bf k},\dot{\chi}_{\bf k})\vert_{\eta_0}$ with $\mu=3/2$  [see Eqs. (\ref{BD},\ref{dchi})]. Using relations (\ref{dotR},\ref{dotS}) and the asymptotics for the functions $P^{3/2}_\nu$ and  $Q^{3/2}_\nu$, one can check (see Appendix \ref{A2:beta-coef} for further details) that $\bar{\beta}_{\bf k}$ in Eq. \eqref{beta} is of order $O\left(k^{-2}\right)$ in the ultraviolet regime $k\to \infty$ (or, equivalently, when $\nu\to \infty$). This asymptotic behavior is sufficient to satisfy the summability condition, and it is therefore proven that the Fock representation discussed in the previous section is unitarily equivalent to the representation defined by the Allen-Folacci vacua.

In addition to the massless case, which corresponds to $\mu=3/2$, one can further check (as we do in Appendix \ref{A2:beta-coef}) that the asymptotic behavior $\bar{\beta}_{\bf k}\sim O\left(k^{-2}\right)$ still holds for any $m>0$, case in which the solution (\ref{BD}) corresponds to the celebrated Bunch-Davies (or Euclidean) vacuum. Hence, the Fock representation of Sec. \ref{sec:unitary dynamics} is unitarily equivalent to the representation based on the Bunch-Davies vacuum. Let us again remark that the quantization of Sec. \ref{sec:unitary dynamics} supports a unitary dynamics for massive fields as well. We have thus proven, also for massive fields, that the quantization based on the  Bunch-Davies vacuum, which follows from the requirement of full de Sitter invariance and the Hadamard condition, is unitarily equivalent to the quantization obtained from the requirement of unitary implementation of the dynamics.

The fact that both viewpoints --the one using the Hadamard condition and the one based on unitary dynamics-- select the same equivalence class of representations [when the scaled field $\chi$ is used] can only be considered as a reassuring result, connecting two {\em a priori} distinct approaches. In the unitary dynamics perspective, one imposes only the existence of unitary transformations that implement the classical time evolution between any two (regular) instants separated by a finite (as opposed to {\em infinitesimal}) interval of time, with no extra requirements, like e.g. continuity or any pre-assigned local form of the vacuum. When one uses the Hadamard condition, an apparently stronger condition is imposed, that fixes the local singularity structure of the vacuum state. As it is well known, this condition is strong enough to guarantee the regularization of the stress-energy tensor, which was in fact the original motivation to adhere to it. It does not seem at all clear that those two viewpoints
should lead to fully equivalent quantizations, and the fact that they do constitutes an interesting result by itself.

\section{Conclusions and discussion}
\label{section: conclusions}

We have explicitly shown that there exists a Fock quantization of the massless scalar field in de Sitter spacetime admitting a unitary implementation of the time evolution. Like in other situations considered in the literature, this involves a scaling of the field variable, using  the conformal factor, and the introduction of a suitable momentum field, given by the conformal time derivative of the scaled field. Our result disproves previous statements in the literature \cite{VV} claiming precisely that it is impossible to attain a Fock quantization with unitary dynamics in this case, by means of the scaling that has proven to be so effective in other situations \cite{PRD79,CMV8,CMOV-FTC,unit-gt3,unique-gowdy-1,scho,BVV2,CQG25}. In addition to providing a direct proof of the unitary implementation of the dynamics of the massless scalar field in de Sitter spacetime, we have analyzed a completely solvable toy model, which further clarifies the viability of a unitary evolution.

Besides, we have seen that the same Fock representation also supports a unitary implementation of the dynamics of massive free fields. It is worth remarking that general results allow us to ensure that the Fock representation depicted in Sec. \ref{sec:unitary dynamics} is the unique one (modulo unitary equivalence) admitting a unitary implementation of the time evolution of free fields in de Sitter spacetime. Fortunately, there is no tension between this result, on the one hand, and the uniqueness provided by imposing the Hadamard criterion, on the other hand. Let us recall that, for free scalar fields propagating on spatially compact universes, the Hadamard approach selects a unique preferred representation of the canonical commutation relations \cite{wald}. Thus, in particular, a Klein-Gordon field $\phi$ in de Sitter spacetime has a unique quantum Hadamard representation. More specifically, given a massive free field in de Sitter spacetime, there exists a unique $O(1,4)$-invariant Fock vacuum state
satisfying the Hadamard condition: the Bunch-Davies (or Euclidean) vacuum state. In the massless case, there are instead infinitely many $O(4)$-invariant Hadamard vacua, differing in their particular zero mode sector parametrization \cite{AllenFol}. However, since the discrepancy between vacua involve just a finite number of degrees of freedom, the Stone-von Neumann uniqueness theorem guarantees that the family of $O(4)$-invariant Hadamard vacua is, in fact, a family of unitarily equivalent states. Thus, in de Sitter spacetime, there is a unique $O(1,4)$-invariant Hadamard quantization of a massive Klein-Gordon field, and a unique (equivalence class of) $O(4)$-invariant Hadamard quantization(s) of the massless Klein-Gordon field. Under the time dependent canonical transformation \eqref{tdct}, the unique Hadamard quantization of the ($m\geq0$) Klein-Gordon field $\phi$ determines, in the scaled field description $\chi$, a quantum
theory which is characterized  by the Cauchy data (\ref{BD},\ref{dchi}). Our results of Sec. \ref{sec:hadamard equivalence} show that for any (nonnegative) value of the mass parameter, this translated (unique) Hadamard quantum theory defines a Fock quantization which is unitarily equivalent to our Fock representation; i.e., the so translated Hadamard quantization and the Fock representation with unitary dynamics provide exactly the same physical predictions. This equivalence eliminates any concern about a possible tension between the requirement of a unitary time evolution (together with the invariance under the spatial symmetries) and the Hadamard condition in order to select a Fock representation.

\section*{Acknowledgements}

We acknowledge financial support from the research grants MICINN/MINECO FIS2011-30145-C03-02 from Spain, CERN/FP/116373/2010 from Portugal, and DGAPA-UNAM IN117012-3 from Mexico. D. M-dB was supported by CSIC and the European Social Fund under the grant JAEPre\_09\_01796. D. M-dB also acknowledges the Faculdade de Ci\^{e}ncias da Universidade da Beira Interior for its warm hospitality during the preparation of part of this work.

\appendix

\section{Equivalence with adiabatic states}

\label{section: adiabatic states}
\label{A1:Adiab-states}

For the sake of completeness, we devote this appendix to discuss the relationship between our Fock quantization with unitary dynamics and the use of adiabatic states in de Sitter spacetime. As it is known, adiabatic states were introduced in the late sixties by Parker \cite{Parker} to bring forward the best possible definition of {\em particles} in expanding universes scenarios. In the framework of closed universes (specifically, in FRW cosmologies with $k=+1$) two important mathematical-physics results exist concerning adiabatic states: (i) the family of adiabatic vacua is a set of unitarily equivalent states \cite{LR}, and (ii) adiabatic states are unitarily equivalent to Hadamard states (for $m>0$) \cite{Junker}. We thus have, in particular, that the (Hadamard) Bunch-Davies vacuum state is unitarily equivalent to an adiabatic state. Given the equivalence between the Bunch-Davies state and the unitary Fock vacuum state (namely, the vacuum of our Fock quantization with unitary time evolution),
established in Sec. \ref{sec:hadamard equivalence}, we then conclude that our
Fock vacuum is unitarily equivalent to an adiabatic state. We will explicitly show here that our Fock state is unitarily equivalent to the zeroth order adiabatic vacuum state.

In order to properly compare the zeroth order adiabatic vacuum state, defined in the $\phi$-description, with the unitary Fock vacuum state, defined in the scaled $\chi$-description, we will proceed in three steps. First, we will choose $\chi$ as the basic field variable. Next, we will translate the Cauchy data defining the zeroth order adiabatic state to the $\chi$-description, and finally we will compare the result with our unitary Fock vacuum state, defined by the Cauchy data \eqref{ourdata}.

Let us start by considering the field equation \eqref{infeq} for a free scalar field propagating in de Sitter spacetime. By performing a mode decomposition of the field, we get that the time-dependent part of the mode solutions, $v_{\bf k}$, obey the second-order differential equation
\begin{equation}
 \label{ineom}
\partial^{2}_{t} v_{\bf k}+3\left(\frac{\partial_{t}a}{a}\right)\partial_{t}v_{\bf k} + w^{2}_{k}v_{\bf k} =0, \qquad w^{2}_{k}=\frac{k(k+2)}{a^{2}}+m^{2}.
\end{equation}
The modes $v_{\bf k}$ must satisfy, in addition, the normalization condition
\begin{equation}
  \label{useT}
 Q_{\bf k}P^{\ast}_{\bf k} - Q_{\bf k}^{\ast}P_{\bf k}=i, \qquad \forall {\bf k},
\end{equation}
where $Q_{\bf k}=v_{\bf k}(t_{0})$ and $P_{\bf k}=a^{3}\partial_{t}v_{\bf k}(t_0)$ are, respectively, the configuration and momentum coefficients of the field $\phi$ on the Cauchy surface $t=t_0$.

In order to introduce adiabatic vacuum states, let us consider solutions to Eq. \eqref{ineom} of the form
\begin{equation}
\label{u-proposal}
 v_{\bf k}(t)=\frac{1}{\sqrt{2a^3\Theta_{\bf k}}}\exp{\left(-i \int_{\bar{t}}^{t}\Theta_{\bf k} (\tilde{t})d\tilde{t}\right)},
\end{equation}
where $\Theta_{\bf k}$ are real positive functions which, according to Eq. \eqref{ineom} and Eq. \eqref{u-proposal}, must satisfy
\begin{equation}
 \label{Teom}
 \Theta_{\bf k}^{2}=w_{k}^{2}-\frac{3}{4}\left(\frac{\partial_{t}a}{a}\right)^{2}-\frac{3}{2}\frac{\partial_{t}^{2}a}{a}+\frac{3}{4}\left(\frac{\partial_{t}\Theta_{\bf k}}{\Theta_{\bf k}}\right)^{2}-\frac{1}{2}\frac{\partial_{t}^{2}\Theta_{\bf k}}{\Theta_{\bf k}}.
\end{equation}
This equation can be solved via an iterative process whenever a finite time interval and a sufficiently large $k$ are considered \cite{LR}. Thus, starting the process with $\Theta^{(0)}_{\bf k}=w_{k}$, we get in the left-hand side of Eq. \eqref{Teom} the $(r+1)$-th function $\Theta^{(r+1)}_{\bf k}$ by plugging in the right-hand side the preceding $r$-th solution $\Theta^{(r)}_{\bf k}$.

An adiabatic vacuum state of $r$-th order is a Fock state constructed from a solution $\{v_{\bf k}\}$ to Eq. \eqref{ineom} with initial conditions $\big(v^{(r)}_{\bf k}(t_{0}),\partial_{t}v^{(r)}_{\bf k}(t_{0})\big)$, where{\footnote{It is worth remarking that adiabatic vacuum states are independent of the values chosen for $\bar{t}$ and $t_0$. Indeed, different choices of $\bar{t}$ in Eq. \eqref{w-rth-order} just introduce  irrelevant phases, whereas distinct elections of the reference (initial) time $t_0$ lead to equivalent vacuum states \cite{LR}.}}

\begin{equation}
\label{w-rth-order}
v^{(r)}_{\bf k}(t_0)=\frac{1}{\sqrt{2a^3\Theta^{(r)}_{\bf k}}}\exp{\left(-i \int_{\bar{t}}^{t_0}\Theta^{(r)}_{\bf k} (\tilde{t})d\tilde{t}\right)}.
\end{equation}

We have $\Theta^{(0)}_{\bf k}=w_{\bf k}=\left[k(k+2)+(ma)^{2}\right]^{1/2}/a$ for the zeroth order, so that the Cauchy data of the corresponding adiabatic vacuum state is given by
\begin{equation}
\label{0th-cd-phi}
Q_{\bf k}=v^{(0)}_{\bf k}, \qquad P_{\bf k}=-a^{2}v^{(0)}_{\bf k}\left[\left(1+\frac{m^{2}}{2w^{2}_{k}}\right)\partial_{t}a+iaw_{k}\right].
\end{equation}
Now, we translate the Cauchy data \eqref{0th-cd-phi} to the $\chi$-description via the time-dependent canonical transformation \eqref{tdct},
\begin{equation}
\label{cd0thoadvschi}
q_{\bf k}=av^{(0)}_{\bf k}\Big\vert_{\eta_{0}}, \qquad p_{\bf k}=-\frac{1}{2}v^{(0)}_{\bf k}\left[ \dot{a} \frac{m^{2}}{w^{2}_{k}}+2ia^{2}w_{k}\right]\Big\vert_{\eta_{0}},
\end{equation}
where $\eta_{0}=2\arctan(e^{Ht_{0}})$ [see Eq. \eqref{eta-t-relation}].

Recall that the Cauchy data defining the unitary Fock vacuum state are given by Eq. \eqref{ourdata},
$\big(u_{\bf k}(\eta_0)=(2\omega_{k})^{-1/2}, \dot{u}_{\bf k}(\eta_0)=-i(\omega_{k}/2)^{1/2}\big)$. Thus, the antilinear part of the Bogoliubov transformation relating the Cauchy data (\ref{ourdata}) and (\ref{cd0thoadvschi}) reads
\begin{equation}
\tilde{\beta}_{\bf k}=\frac{1}{\sqrt{2\omega_{k}}}\left(\omega_{k}q_{\bf k}-ip_{\bf k}\right)=\frac{e^{-i\int w_{ k}}}{2(1-x^{2}_{k})^{1/4}}\left[i\frac{\partial_{t}a}{2ma^{2}}x_{k}^{3}-\left(1-\sqrt{1-x_{k}^{2}}\right)\right],
\end{equation}
where $x_{k}=m/w_{k}$. The unitary Fock vacuum and the adiabatic vacuum will be unitarily equivalent states if and only if $\tilde{\beta}_{\bf k}$ is square summable. It is a simple exercise to see that, in the asymptotic regime, which corresponds to $x_{k}<<1$, the ultraviolet behavior of beta is $\tilde{\beta}_{\bf k}\sim O(k^{-2})$. As a consequence, we conclude that the unitary Fock vacuum state is equivalent to the zeroth order adiabatic vacuum. Now, since in closed FRW spacetimes any two adiabatic states of distinct order are unitarily equivalent, we have in fact that our equivalence result extends to adiabatic states of arbitrary order.

\section{Ultraviolet behavior of the beta functions and coefficients}
\label{A2:beta-coef}

In this appendix, we will detail the derivation of the ultraviolet behavior, i.e., the behavior for large $k$, of the beta functions $\beta_{\bf k}(\eta_{0},\eta)$ of the evolution transformation, defined in Sec. \ref{sec:unitary dynamics}, and of the beta coefficients $\bar{\beta}_{\bf k}$ of the canonical transformation that relates the vacuum selected by the unitary evolution criterion and the Allen-Folacci vacuum, discussed in Sec. \ref{sec:hadamard equivalence}.

Let us start by detailing the behavior for large $k$ of the beta functions, i.e., the antilinear part of the evolution transformation $\mathcal{U}_{\bf k}(\eta_{0},\eta)$ introduced in Eq. \eqref{evolution-a}. We first recall the expression of this transformation on the creation and annihilation variables in terms of the evolution of the configuration and momentum modes,
\begin{equation}
 \mathcal{U}_{\bf k}(\eta_{0},\eta)=T_{k}U_{k}(\eta_0,\eta)T^{-1}_k=T_{k}M_{k}(\eta)M^{-1}_{k}(\eta_0)T^{-1}_k.
\end{equation}
The matrices $M_{k}$ and $T_{k}$ are given in Eq. \eqref{defM} and
Eq. \eqref{defT}, respectively. It follows from this expression that the beta functions are
\begin{equation}
\label{betafunexp}
\beta_{\bf k}(\eta_{0},\eta)=\frac{\Gamma\left(\frac{\nu-\mu+2}{2}\right)\Gamma\left(\frac{\nu-\mu+1}{2}\right)}{2^{{2\mu+1}}\Gamma\left(\frac{\nu+\mu+2}{2}\right)\Gamma\left(\frac{\nu+\mu+1}{2}\right)}\left[\Delta R^{\mu}_{\nu}\dot{S}^{\mu}_{\nu}-\Delta S^{\mu}_{\nu}\dot{R}^{\mu}_{\nu}+i\frac{\Delta \dot{R}^{\mu}_{\nu}\dot{S}^{\mu}_{\nu}}{\omega_{k}}+i\omega_{k}\Delta S^{\mu}_{\nu}R^{\mu}_{\nu} \right],
\end{equation}
where we have introduce the notation
\begin{equation}
\Delta F^{\mu}_{\nu}G^{\mu}_{\nu}=F^{\mu}_{\nu}(-\cos\eta)G^{\mu}_{\nu}(-\cos\eta_{0})-G^{\mu}_{\nu}(-\cos\eta)F^{\mu}_{\nu}(-\cos\eta_{0}),
\end{equation}
for any function $F^{\mu}_{\nu}$ and $G^{\mu}_{\nu}$ in $\{R^{\mu}_{\nu},S^{\mu}_{\nu},\dot{R}^{\mu}_{\nu},\dot{S}^{\mu}_{\nu}\}$. Besides, to compute the inverse matrix $M_{k}^{-1}$, we have employed the determinant of $M_{k}$:
\begin{equation}
\det M_{k}(\eta)=\sin^{2}\!\eta\, \mathcal{W}\{P^{\mu}_{\nu}(-\cos\eta),Q^{\mu}_{\nu}(-\cos\eta)\}=\frac{2^{2\mu}\Gamma\left(\frac{\nu+\mu+2}{2}\right)\Gamma\left(\frac{\nu+\mu+1}{2}\right)}{\Gamma\left(\frac{\nu-\mu+2}{2}\right)\Gamma\left(\frac{\nu-\mu+1}{2}\right)}.
\end{equation}
Here, $\mathcal{W}\{\cdot,\cdot\}$ denotes the Wronskian.
By considering the relations between the functions $P^{\mu}_{\nu}(x)$ and $Q^{\mu}_{\nu}(x)$ with $P^{\mu}_{\nu}(-x)$ and $Q^{\mu}_{\nu}(-x)$, given in expressions (\ref{minus1},\ref{minus2}), one gets
\begin{eqnarray}
\Delta R^{\mu}_{\nu}\dot{S}^{\mu}_{\nu}=\dot{S}^{\mu}_{\nu}(\cos\eta)R^{\mu}_{\nu}(\cos\eta_{0})-R^{\mu}_{\nu}(\cos\eta)\dot{S}^{\mu}_{\nu}(\cos\eta_{0}),
\end{eqnarray}
and similarly for {$\Delta S^{\mu}_{\nu}\dot{R}^{\mu}_{\nu}$}, $\Delta \dot{R}^{\mu}_{\nu}\dot{S}^{\mu}_{\nu}$, and $\Delta S^{\mu}_{\nu}R^{\mu}_{\nu}$.

Using then Eqs. (\ref{seexR},\ref{seexS}) for the asymptotic expansion of the associated Legendre functions, as well as the definition \eqref{defRS} of the functions $R^{\mu}_{\nu}$ and $S^{\mu}_{\nu}$, and the expressions (\ref{dotR},\ref{dotS}) of their time derivatives, one finds that the beta functions have the asymptotic behavior
\begin{equation}
\label{betaexp1}
\beta_{\bf k}(\eta,\eta_{0})=-\frac{\Gamma\left({\frac{2k-2\mu+5}{4}}\right)\Gamma\left({\frac{2k-2\mu+3}{4}}\right)}{2^{{2\mu+1}}\Gamma\left({\frac{2k+2\mu+5}{4}}\right)
\Gamma\left({\frac{2k+2\mu+3}{4}}\right)}
\frac{\Gamma^{2}\left(k+\mu+3/2\right)}{\Gamma\left(k+1\right)\Gamma\left(k+2\right)}\sum_{j=0}\mathfrak{B}_{j}(\eta,\eta_{0}),
\end{equation}
where the functions $\mathfrak{B}_{j}(\eta,\eta_{0})$ are of asymptotic order $O\left(\omega_{k}^{-j}\right)$. Employing the Stirling formula for the asymptotic behavior of the Gamma function \cite{AS},
\begin{equation}
\label{stirl}
\Gamma(z+1) \sim \sqrt{2\pi z}\left(z/e\right)^{z},
\end{equation}
it follows that the time-independent coefficient that multiplies the functions $\mathfrak{B}_{j}(\eta,\eta_{0})$ in Eq. \eqref{betaexp1} behaves asymptotically as $O(1)$. Therefore, the beta functions have the same ultraviolet behavior as the first nonvanishing function $\mathfrak{B}_{j}(\eta,\eta_{0})$. We will explicitly show that both $\mathfrak{B}_{0}(\eta,\eta_{0})$ and $\mathfrak{B}_{1}(\eta,\eta_{0})$ vanish, proving that the beta function is (at least) of the asymptotic order $O\left(\omega^{-2}_{k}\right)\sim O\left(k^{-2}\right)$.

The function $\mathfrak{B}_{0}(\eta,\eta_{0})$ can be deduced from Eq. \eqref{betafunexp} by considering only the leading asymptotic contributions of the terms of the form $\Delta F^{\mu}_{\nu}G^{\mu}_{\nu}$ in that expression. For this, one can use the asymptotic expansions of $R^{\mu}_{\nu}$, $S^{\mu}_{\nu}$, and their time derivatives, which can be obtained from Eqs. (\ref{seexR}) and (\ref{seexS}) as well as Eqs. (\ref{dotR}) and (\ref{dotS}). With a bit of calculus, and using trigonometric relations, one gets the following four leading contributions, which annihilate each other in pairs:
\begin{eqnarray}
\mathfrak{B}_{0}(\eta,\eta_{0})= &-\sin\left[(k+1)(\eta{+}\eta_{0})+\mu\pi\right]+\sin\left[(k+1)(\eta{+}\eta_{0})+\mu\pi\right] \qquad \qquad \nonumber \\ &+i\cos\left[[(k+1)(\eta{+}\eta_{0})+\mu\pi\right]-i\cos\left[[(k+1)(\eta{+}\eta_{0})+\mu\pi\right] = 0.
\end{eqnarray}

On the other hand, the expression of the function $\mathfrak{B}_{1}(\eta,\eta_{0})$ can be computed in the same way, but considering now the next leading contributions of the terms of the type $\Delta F^{\mu}_{\nu}G^{\mu}_{\nu}$. These are obtained from the products of the leading and first-subleading contributions in the asymptotic expansions of $R^{\mu}_{\nu}$, $S^{\mu}_{\nu}$, and of their time derivatives. As before, making use of trigonometric relations, one obtains
\begin{eqnarray}
\frac{8(k+1)}{4\mu^{2}-1}\mathfrak{B}_{1}(\eta,\eta_{0})\!&=\!& \displaystyle\frac{1}{\sin\eta}\left\{\cos\left[(k+2)\eta+(k+1)\eta_{0}+\mu\pi\right]-\cos\left[(k+2)\eta+(k+1)\eta_{0}+\mu\pi\right]\right. \qquad \quad  \nonumber\\
& & -\left. i\sin\left[(k+2)\eta+(k+1)\eta_{0}+\mu\pi\right]+i\sin\left[(k+2)\eta+(k+1)\eta_{0}+\mu\pi\right]\right\} \nonumber\\
& & \displaystyle+\frac{1}{\sin\eta_{0}}\left\{\cos\left[(k+1)\eta+(k+2)\eta_{0}+\mu\pi\right]-\cos\left[(k+1)\eta+(k+2)\eta_{0}+\mu\pi\right]\right. \nonumber\\
& & -\left. i\sin\left[(k+1)\eta+(k+2)\eta_{0}+\mu\pi\right]+i\sin\left[(k+1)\eta+(k+2)\eta_{0}+\mu\pi\right]\right\} = 0.
\end{eqnarray}
Therefore, both $\mathfrak{B}_{0}(\eta,\eta_{0})$ and $\mathfrak{B}_{1}(\eta,\eta_{0})$ vanish and, consequently, the asymptotic behavior of the beta functions is given by $\mathfrak{B}_{2}(\eta,\eta_{0})$ (assuming that it does not vanish). Hence, $\beta(\eta,\eta_{0})\sim O\left(k^{-2}\right)$, as we wanted to prove. Note that this result is valid for any value of $\mu$ and thus holds for every possible mass of the scalar field. In particular, it is so in the massless case, corresponding to $\mu=3/2$.

Let us now study the beta coefficients $\bar{\beta}_{\bf k}$ of the transformation that relates the Allen-Folacci vacua and the vacuum that defines the Fock representation with unitary evolution. Recalling the definition \eqref{beta} of these beta coefficients, as well as the initial data that define both types of vacua [see expressions \eqref{ourdata}, \eqref{BD}, and \eqref{dchi}], one arrives at
\begin{equation}
\bar{\beta}_{\bf k}=-A_{k}\frac{e^{i(k+\mu)\pi}}{\sqrt{2{\omega_{k}}}}\left[
\frac{2\omega_{k}}{\pi}S^{\mu}_{\nu}(\cos\eta_{0})-\dot{R}^{\mu}_{\nu}(\cos\eta_{0})
-i\frac{2}{\pi}\dot{S}^{\mu}_{\nu}(\cos\eta_{0})-i{\omega_{k}}R^{\mu}_{\nu}(\cos\eta_{0})\right].
\end{equation}
As in previous calculations, we introduce the expressions of the functions $R^{\mu}_{\nu}$, $S^{\mu}_{\nu}$, and their time derivatives, in terms of the associated Legendre functions. Then, from the asymptotic expansions (\ref{seexR},\ref{seexS}) of those functions, and using Eq. \eqref{ABeucvac}, one finds that the beta coefficients admit an asymptotic expansion of the form
\begin{equation}
\bar{\beta}_{\bf k}=-\sqrt{\frac{\Gamma(k-\mu+3/2)}{\Gamma(k+\mu+3/2)}}\frac{e^{i(k+3\mu/2)\pi}}{2\sqrt{\omega_{k}}}
\frac{\Gamma(k+\mu+3/2)}{\Gamma(k+1)}\sum_{j=0}\bar{\mathfrak{B}}_j.
\end{equation}
Here, $\bar{\mathfrak{B}}_j$ are functions of the asymptotic order $O\left(\omega^{-j}\right)\sim O\left(k^{-j}\right)$. Employing again the Stirling formula \eqref{stirl}, it follows that the global coefficient appearing in front of the sum of functions $\bar{\mathfrak{B}}_{j}$ is of order ${O(1)}$. Therefore, as in the previous case, the asymptotic order of the beta coefficients will coincide with that of the first nonvanishing contribution $\bar{\mathfrak{B}}_{j}$. In this case, the function $\bar{\mathfrak{B}}_{0}$ reads
\begin{eqnarray}
\bar{\mathfrak{B}}_{0}&=&\cos\left[(k+1)\eta_{0}+\frac{\pi}{4}+\frac{\mu\pi}{2}\right]+\sin\left[(k+1)\eta_{0}-\frac{\pi}{4}+\frac{\mu\pi}{2}\right]\nonumber \\ & & +i\sin\left[(k+1)\eta_{0}+\frac{\pi}{4}+\frac{\mu\pi}{2}\right]-i\cos\left[(k+1)\eta_{0}-\frac{\pi}{4}+\frac{\mu\pi}{2}\right]=0.
\end{eqnarray}
On the other hand, one can check that the function $\bar{\mathfrak{B}}_{1}$ is given by
\begin{eqnarray}
\bar{\mathfrak{B}}_{1}&=&\frac{4\mu^{2}-1}{8(k+2)\sin\eta_{0}}\left\{-\cos\left[(k+2)\eta_{0}-\frac{\pi}{4}+\frac{\mu\pi}{2}\right]+\sin\left[(k+2)\eta_{0}+\frac{\pi}{4}+\frac{\mu\pi}{2}\right]\right.\nonumber  \qquad \qquad \\ & & \qquad \qquad  -i\left.\sin\left[(k+2)\eta_{0}-\frac{\pi}{4}+\frac{\mu\pi}{2}\right]-i\cos\left[(k+2)\eta_{0}+\frac{\pi}{4}+\frac{\mu\pi}{2}\right]
\right\}=0.
\end{eqnarray}
Therefore, as we wanted to show, the asymptotic behavior of the beta coefficients is at most of the order $\bar{\mathfrak{B}}_{2}\sim O\left(k^{-2}\right)$. This guarantees their square summability (counting the degeneracy), and hence the unitary implementation of the transformation in the considered Fock representation. Again, this result is independent of the value of $\mu$, and therefore is valid not only for the massless case ($\mu=3/2$), but also for an arbitrary mass of the scalar field.

\bibliographystyle{plain}

\end{document}